\documentclass{aa}
\usepackage{graphicx}
\usepackage{txfonts}
\begin{document}

\def\pc{{\rm\thinspace pc}}
\def\atpcmcube  {{\rm\thinspace atoms~cm^{-3}}}
\def\pcm3{{\rm\thinspace cm^{-3}}}
\def\pk{{\rm\thinspace \rm cm^{-3} \thinspace K}}
\def\pcmsq{{\rm\thinspace cm^{-2}}}
\def\lI{{\rm Loop I\thinspace}}
\def\contcaption{\@conttrue\SFB@caption\@captype}
\newcommand{\ic}{\mbox {$\rm I_{C} $}}
\newcommand{\mk}{\mbox {$\rm K_{MK} $}}
\newcommand{\msun}{\mbox {$\rm M_{\odot} $}}
\newcommand{\mjup}{\mbox {$\rm M_{J} $}}
\newcommand{\teff}{\mbox {$\rm T_{eff} $}}
\newcommand{\cmccc}{\mbox{$\rm cm^{-6}~pc $}}
\newcommand{\hi}{\mbox{$\rm {H\,{\sc i}}\:$}}
\newcommand{\hii}{\mbox{$\rm {H\,{\sc ii}}\:$}}
\newcommand{\hei}{\mbox{$\rm {He\,{\sc i}}\:$}}
\newcommand{\heii}{\mbox{$\rm {He\,{\sc ii}}\:$}}
\newcommand{\heiii}{\mbox{$\rm {He\,{\sc iii}}\:$}}
\newcommand{\civ}{\mbox{$\rm {C\,{\sc iv}}\:$}}
\newcommand{\siiv}{\mbox{$\rm {Si\,{\sc iv}}\:$}}
\newcommand{\ovi}{\mbox{$\rm {O\,{\sc vi}}\:$}}
\newcommand{\ciii}{\mbox{$\rm {C\,{\sc iii}}\:$}}
\newcommand{\cii}{\mbox{$\rm {C\,{\sc ii}}\:$}}
\newcommand{\kms}{\mbox{$\rm km\thinspace\s^{-1}\:$}}
\newcommand{\cms}{\mbox{$\rm cm^{-2}\:$}}
\newcommand{\cmc}{\mbox{$\rm cm^{-3}\:$}}
\def\n_h{{\rm n_{H}}}
\newcommand{\fhe}{\mbox{$\rm f_{He}\:$}}
\newcommand{\nh}{\mbox{$\rm N_{H}\:$}}
\newcommand{\vhii}{\mbox{$\rm n_{HII}\:$}}
\newcommand{\nhii}{\mbox{$\rm N_{H\,{\sc ii}}\:$}}
\newcommand{\nhi}{\mbox{$\rm N_{H\,{\sc i}}\:$}}
\newcommand{\nhei}{\mbox{$\rm N_{He\,{\sc i}}\:$}}
\newcommand{\nheii}{\mbox{$\rm N_{He\,{\sc ii}}\:$}}
\newcommand{\nheiii}{\mbox{$\rm N_{He\,{\sc iii}}\:$}}
\def\ism{{\rm interstellar medium }}
\def\lism{{\rm local interstellar medium }}
\def\ts{{\rm time-slot }}
\def\tl{{\sc TLUSTY}}
\def\syn{{\sc SYNSPEC}}
\def\euve{{\sc EUVE}}
\def\axaf{{\sc AXAF}}
\def\fuse{{\sc FUSE}} 
\def\WFC{{\sl WFC }}
\def\PSPC{{\sl PSPC }}
\def\ROSAT{{\sl ROSAT }}
\def\ne{{n_{\rm e}~}}
\def\hy{{\rm H }}
\def\he{{\rm He }}
\def\nhd{{$n_{\rm H}~$}}
\def\nhyi{{$n_{\rm HI}~$}}
\def\NH1{{$N_{\rm HI}~$}}
\def\nho{{$n_{\rm o}~$}}
\def\lts{late type stars }
\def\wds{white dwarfs }
\def\Ksec{{\rm\thinspace Ksec }}
\def\ex{{\it EXOSAT }}
\def\gi{{\it Ginga }}
\def\ein{{\sl Einstein }}
\def\lx{{$\rm L_{x}$~(2--10~keV) \ }}
\def\sx{{$\rm S_{x}$~(2--10~keV) \ }}
\def\nhf{{$\rm N_{HFe}$}}
\def\mh{{$\rm M_{H}$}}
\def\fy{{$\rm \omega_{K}$}}
\def\s{\ \ \ }          
\def\ss{\s\ }           
\def\sss{\s\ \ }        
\def\ssss{\s\s}         

\def\ih{{\rm I$_{\mathrm{Harris}}$}}
\def\zr{{\rm Z$_{\mathrm{RGO}}$}}
\def\is{{\rm I$_{\mathrm{SDSS}}$}}
\def\zs{{\rm Z$_{\mathrm{SDSS}}$}}
\def\cm{{\rm\thinspace cm}}
\def\erg{{\rm\thinspace erg}}
\def\eV{{\rm\thinspace eV}}
\def\g{{\rm\thinspace g}}
\def\ga{{\rm\thinspace gauss}}
\def\K{{\rm\thinspace K}}
\def\keV{{\rm\thinspace keV}}
\def\km{{\rm\thinspace km}}
\def\Kpc{{\rm\thinspace kpc}}
\def\Lsun{\hbox{$\rm\thinspace L_{\odot}$}}
\def\Rsun{\hbox{$\rm\thinspace R_{\odot}$}}
\def\m{{\rm\thinspace m}}
\def\Mpc{{\rm\thinspace Mpc~}}
\def\Msun{\hbox{$\rm\thinspace M_{\odot}$}}
\def\mic{{\rm\thinspace $\mu$m}}
\def\ph{{\rm\thinspace ph}}
\def\s{{\rm\thinspace s}}
\def\yr{{\rm\thinspace yr}}
\def\sr{{\rm\thinspace sr}}
\def\Hz{{\rm\thinspace Hz}}
\def\cmps{\hbox{$\cm\s^{-1}\,$}}
\def\pcmsq{\hbox{$\cm^{-2}\,$}}
\def\cmsq{\hbox{$\cm^2\,$}}
\def\ergcmcups{\hbox{$\erg\cm^3\ps\,$}}
\def\ergpcmps{\hbox{$\erg\cm^{-3}\s^{-1}\,$}}
\def\gpcm{\hbox{$\g\cm^{-3}\,$}}
\def\gpcmps{\hbox{$\g\cm^{-3}\s^{-1}\,$}}
\def\gps{\hbox{$\g\s^{-1}\,$}}
\def\kmps{\hbox{$\km\s^{-1}\,$}}
\def\Lsunppc{\hbox{$\Lsun\pc^{-3}\,$}}
\def\Msunpc{\hbox{$\Msun\pc^{-3}\,$}}
\def\Msunpkpc{\hbox{$\Msun\kpc^{-1}\,$}}
\def\Msunppc{\hbox{$\Msun\pc^{-3}\,$}}
\def\Msunppcpyr{\hbox{$\Msun\pc^{-3}\yr^{-1}\,$}}
\def\Msunpyr{\hbox{$\Msun\yr^{-1}\,$}}
\def\pcmc{\hbox{$\cm^{-3}\,$}}
\def\pcmK{\hbox{$\cm^{-3}\K$}}
\def\phpcmsqps{\hbox{$\ph\cm^{-2}\s^{-1}\,$}}
\def\phpcmsqpspkev{\hbox{$\ph\cm^{-2}\s^{-1}\keV^{-1}\,$}}
\def\pHz{\hbox{$\Hz^{-1}\,$}}
\def\ps{\hbox{$\s^{-1}\,$}}
\def\psqcm{\hbox{$\cm^{-2}\,$}}
\def\psr{\hbox{$\sr^{-1}\,$}}
\def\pyr{\hbox{$\yr^{-1}\,$}}


\def\ek{EK UMa }
\def\re{RE1938-461 }
\def\ae{AE Aqr }
\def\sw{SW UMa }
\def\ho{1H0709-360 }
\def\ru{RU Peg }
\def\qs{QS Tel }
\def\chisq{{$\chi^{2}$}}
\def\rchi{{$\chi^{2}_{\nu}$}}
\def\etal{{\it et al.\thinspace}}
\def\eg{{\it e.g.\ }}
\def\etc{{\it etc.\ }}
\def\ie{{\it i.e.\ }}
\def\nhunit{{\rm\thinspace atoms~cm^{-2}}}
\def\atpcm{{\rm\thinspace atoms~cm^{-2}}}
\def\mdot{{$\rm \dot{M}$}}
\def\approxlt{\mathrel{\hbox{\rlap{\lower .5ex \hbox {$\sim$}}
        \raise .15 ex \hbox{$<$}}}}
\def\approxgt{\mathrel{\hbox{\rlap{\lower .5ex \hbox {$\sim$}}
        \raise .15 ex \hbox{$>$}}}}
\def\flux{{\erg\cm^{-2}\s^{-1}}}
\def\ergps{\hbox{$\erg\s^{-1}\,$}}
\def\civ{{C\thinspace IV}}
\def\nv{{N\thinspace V}}
\def\siiv{{Si\thinspace IV}}
\def\la{\mathrel{\hbox{\rlap{\hbox{\lower4pt\hbox{$\sim$}}}\hbox{$<$}}}}
\def\ga{\mathrel{\hbox{\rlap{\hbox{\lower4pt\hbox{$\sim$}}}\hbox{$>$}}}}

\newbox\grsign \setbox\grsign=\hbox{$>$} \newdimen\grdimen
\grdimen=\ht\grsign
\newbox\simlessbox \newbox\simgreatbox \newbox\simpropbox
\setbox\simgreatbox=\hbox{\raise.5ex\hbox{$>$}\llap
     {\lower.5ex\hbox{$\sim$}}}\ht1=\grdimen\dp1=0pt
\setbox\simlessbox=\hbox{\raise.5ex\hbox{$<$}\llap
     {\lower.5ex\hbox{$\sim$}}}\ht2=\grdimen\dp2=0pt
\setbox\simpropbox=\hbox{\raise.5ex\hbox{$\propto$}\llap
     {\lower.5ex\hbox{$\sim$}}}\ht2=\grdimen\dp2=0pt
\def\simgreat{\mathrel{\copy\simgreatbox}}
\def\simless{\mathrel{\copy\simlessbox}}
\def\simprop{\mathrel{\copy\simpropbox}}

   \headnote{Research Note}
   \title{A near-IR spectrum of the DO white dwarf \object{RE J0503-285}\thanks{Based on observations collected at the 
European Southern Observatory, Chile. ESO No. 072.D-0362}
}

   \subtitle{}

   \author{P.D. Dobbie
          \inst{1}
          \and
          M.R. Burleigh\inst{1}
	  \and
	  A.J.Levan\inst{1}
	  \and
	  M.A.Barstow\inst{1}
	  \and
	  R.Napiwotzki\inst{1}
	  \and
	  I.Hubeny\inst{2}
          }

   \offprints{P. Dobbie}

   \institute{Dept of Physics \& Astronomy, University of Leicester,
              Leicester, LE1 7RH, UK\\
              \email{[pdd; mbu; anl; mab]@star.le.ac.uk}
         \and
             Dept of Astronomy and Steward Observatory, University of Arizona, Tucson, AZ 85721, USA \\
             \email{hubeny@as.arizona.edu}
             }

   \date{Received September 15, 1996; accepted March 16, 1997}

   \abstract{
We present a  near-IR spectroscopic analysis of the intriguing DO white dwarf 
RE J0503-285. The IR spectrum fails to reveal evidence of the presence of a spatially unresolved,
cool, late-type companion. Hence we have placed an approximate limit on the spectral-type 
and mass of a putative companion (later than M8, M$<0.085$M$_{\odot}$). This result rules 
out ongoing interaction between the white dwarf and a close companion with 
M$\ge0.085$M$_{\odot}$ as responsible for the discrepancies between the observed photospheric 
abundances and model predictions. As the possibility remains that there is a cooler companion 
lying beyond the detection threshold of this study we use our modelling to estimate the 
improvement in sensitivity offered by a Spitzer observation.

   \keywords{Stars: white dwarfs -- Stars: abundances -- Techniques: spectroscopic --
 Individual: RE J0503-285
               }
   }

   \maketitle
%

\section{Introduction}

In accord with a theoretical prediction made several decades ago (Schatzman \cite{schatzman58}),
the atmospheres of white dwarfs are observed to be dominated either by hydrogen 
(DAs) or by helium (DO/DBs). Unimpeded, their high surface gravities lead to the 
settling out of heavier elements on timescales of mere days to months (e.g. Dupuis et al. 
\cite{dupuis93}). In the standard theory of single star evolution the distinction 
between the two compositions is made at birth. If a star leaves the AGB during a 
period of quiescent hydrogen shell burning it evolves onto the hydrogen 
dominated cooling channel. Alternatively, if a thermal pulse occurs as the star 
evolves off the AGB, much of the remaining hydrogen is incinerated and nuclear 
processed material from deeper layers (e.g. C and O) is dredged up to the surface
leading to a hydrogen deficient object (e.g. Iben et al. 1983, Herwig et al. \cite{herwig99}). 

\begin{table*}
\begin{minipage}{170mm}
\begin{center}
\caption{Summary details of RE J0503-285, including coordinates and near-IR magnitudes obtained 
from the 2MASS All-Sky Point Source Catalogue. The exposure times used for
 the acquisition of the JH and HK near-IR spectra with the NTT and SOFI are also listed.}
\label{sum1}
\begin{tabular}{ccllcccrr}
\hline
\hline
Identity   & Name & \multicolumn{1}{c}{RA}    &  \multicolumn{1}{c}{Dec}   & J &  H & K$_{\rm S}$ & \multicolumn{2}{c}{t$_{\rm exp}$ (secs)} \\
& & \multicolumn{2}{c}{J2000.0} &  &  & & \multicolumn{1}{c}{JH} &  \multicolumn{1}{c}{HK}  \\
 \hline
WD0501-289 & RE J0503-285    & 05 03 55.51  & -28 54 34.6 & $14.65\pm0.03$  & $14.77\pm0.06$ & $14.95\pm0.14$ &  1080 & 2880  \\
\hline
\end{tabular}
\end{center}
\end{minipage}
\end{table*}

A number of empirical results are broadly supportive of this hypothesis. For 
example, the observed number ratios of hydrogen rich to hydrogen-deficient central 
stars of planetary nebulae (2:1; Mendez \cite{mendez91}) and DAs to DOs at T$_{\rm eff}\simgreat
40000$K (7:1; Fleming et al. \cite{fleming86}) are roughly consistent with theoretical 
expectations (Iben \& Tutukov \cite{iben84}). The observed transformation of the hydrogen 
deficient PG1159 central star of the planetary nebula, Lo4, into a Wolf Rayet 
object and back over a period of months (Werner et al. \cite{werner92}) and similarities in 
the measured abundances of the elements C and O in the atmospheres of Wolf Rayet
(C/He$\sim0.2$ and O/He$\sim0.05$, Koesterke \& Hamann \cite{koesterke97}) and PG1159 stars 
(C/He$\sim0.1-0.6$ and O/He$\sim0.005-0.1$, Dreizler \& Heber \cite{dreizler98}) confirm an 
evolutionary link between these two classes of object. Furthermore, the abundance 
patterns observed in the atmospheres of the DO white dwarfs ($120000 \simgreat 
$T$_{\rm eff}\simgreat45000$K, C/He$\sim0.001-0.01$, O/He$\simless0.001$; Dreizler 
\cite{dreizler99}) suggest they are the descendents of the PG1159 stars ($180000\simgreat 
$T$_{\rm eff}\simgreat65000$K), gravitational settling having reduced 
the abundances in the former. 

Despite detecting $\sim$110 hot DA white dwarfs (T$_{\rm eff}\simgreat 25000$K), the 
extreme-ultraviolet surveys of the ROSAT WFC and EUVE unearthed only one new DO white
dwarf, RE J0503-285. An analysis of IUE and Voyager data and the optical identification 
spectrum revealed an effective temperature and surface gravity of T$_{\rm eff}\approx 
70000$K and log g =7.5-8.0 respectively (Barstow et al. \cite{barstow94a}). These values are 
consistent with more recent determinations utilising more refined synthetic spectra 
and/or higher S/N optical data (T$_{\rm eff}=72660_{-6289}^{+2953}$K, log g 
$=7.50^{+0.13}_{-0.15}$, Barstow et al. \cite{barstow00}; T$_{\rm eff}=70000$K, log g=7.50, 
Dreizler \& Werner \cite{dreizler96}). 

Somewhat counter-intuitively to it's detection as an EUV source, of the 14 DOs studied in 
detail by Dreizler \& Werner (\cite{dreizler96}) and Dreizler (\cite{dreizler99}), RE J0503-318 is one of the most
metal rich, with C/He$\approx0.005$, O/He$\approx0.0005$, N/He$\approx10^{-5}$ and 
Si/He$\approx10^{-5}$. Nickel has also been detected in the photosphere at an abundance 
of Ni/H$\approx10^{-5}$ (Barstow et al. \cite{barstow00}). However, no convincing evidence of the 
presence of iron is found in any of the spectral datasets obtained to date, setting an 
upper limit on its abundance of Fe/H$\simless10^{-6}$. 
To an extent this is an intriguing result when one notes that the cosmic abundance ratio of 
Fe:Ni is 18:1, that Fe is observed to be more abundant than Ni in the atmospheres of hot DA 
white dwarfs by factors between 1 to 20 and that self-consistent model calculations taking 
into account gravitational settling and radiative levitation predict an apparent Fe:Ni ratio 
of $\sim$1:3 (e.g. Dreizler \cite{dreizler99}). Barstow \& Sion (\cite{barstow94b}) report evidence of 
episodic massloss in a series of IUE spectra of RE J0503-285. It is plausible that the
observed pattern of abundances may be reflective of this process. Alternatively, the abundance 
pattern may be related to the hypothetical dredging up of nuclear processed material from 
deeper layers caused by the late-thermal pulse, as has been proposed to explain the Fe 
deficiency observed in a number of PG1159 stars (e.g. Miksa et al. \cite{miksa02}). However, 
as the mass of RE J0503-285 maybe somewhat lower than the canonical DO value of $0.59\pm0.08$M$
_{\odot}$ (Dreizler \& Werner \cite{dreizler96}) it is a possibility that binary evolution 
(e.g. Vennes et al. \cite{vennes98}) has influenced or continues to interfere with
the photospheric composition. Indeed, as discussed by Bragaglia et al. (\cite{brag90}) it is possible that 
close binary evolution produces preferentially non-DA white dwarfs with hydrogen-deficient 
envelopes. 
 
The lack of any prior published detailed examination of this particularly 
interesting object at near-IR wavelengths has motivated us to obtain, during a recent study of DA 
white dwarfs, a JHK spectrum of RE J0503-285 to investigate external factors possibly influencing
 its evolution and photospheric composition e.g. the presence of a cool low mass companion. We 
present here the results of our analysis of this data.

\section{Observations}

\subsection{Data acquisition}

A low resolution near-IR spectrum of RE J0503-285 was obtained using the ESO New Technology 
Telescope (NTT) and the Son-of-Isaac (SOFI) infrared instrument on 2003/12/10. The sky conditions 
at the La Silla site were good on this night with seeing typically in the range 0.6''-1.0''. We 
used the low resolution spectroscopic mode ($\lambda/\delta\lambda\sim950$ with the 0.6'' slit),
in which coverage of the wavelength ranges $0.95-1.64\mu$m and $1.53-2.52\mu$m is provided by 
the ``blue'' and the ``red'' grism respectively. The observations were undertaken using the 
standard technique of nodding our point source target back and forth along the spectrograph 
slit in an ABBA pattern. The total integration times used for the blue and the red grism setup
were 1080s and 2880s respectively. To facilitate the removal of telluric features from the
target spectra and to provide an approximate flux calibration, a standard star (HIP28999) was 
observed immediately after the science integrations. This was carefully chosen to lie within 
$\sim0.1$ airmasses of the white dwarf. 

\subsection{Data reduction}

We have used software routines in the STARLINK packages KAPPA and FIGARO to apply standard 
reduction techniques to our data. Further details of these procedures are given in Dobbie et al.
(\cite{dobbie05}). Here we re-iterate that any features intrinsic to the energy distribution of the standard star 
were identified by reference to a near-IR spectral atlas of fundamental MK standards 
(Wallace et al. \cite{wallace00}, Meyer et al. \cite{meyer98}, Wallace \& Hinkle \cite{wallace97}) and were removed by linearly
interpolating over them.  Furthermore, the flux levels of the data were scaled to (1) achieve
the best possible agreement between the blue and the red spectrum of the white dwarf in the 
overlap region between $1.53-1.64\mu$m and (2) obtain the best possible agreement between 
the spectral data and the J, H and K$_{\rm S}$ photometric fluxes for RE J0503-285 derived
from the 2MASS All Sky Data Release Point Source Catalogue magnitudes (Skrutskie et al. \cite{skrutskie97}) 
where zero magnitude fluxes were taken from Zombeck (\cite{zombeck90}). The reduced spectrum and 2MASS 
fluxes are shown in Figure 1.

   \begin{figure*}
   \centering
   \includegraphics{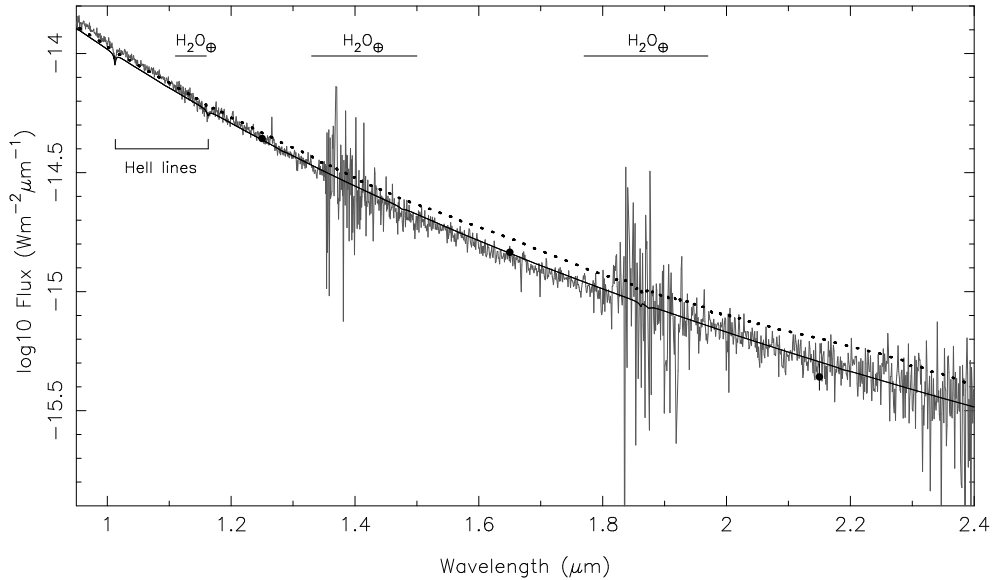}
   \caption{Near-IR spectroscopy (solid grey line) and 2MASS JHK photometry (filled circles) of the white
 dwarf RE J0503-285. A He+C+H non-LTE synthetic white dwarf spectrum of appropriate effective temperature, 
surface gravity and normalisation (solid black line) and a hybrid white dwarf + late-type dwarf model 
representing our estimated limit on the spectral type of a putative spatially unresolved companion 
(dotted black line, WD+M8) are overplotted. We have labelled the more prominent white dwarf HeII lines 
and telluric water vapour features present in this dataset.}
              \label{re0503ir}%
    \end{figure*}

\section{Analysis of the data} 

\subsection{Model DO white dwarf spectra}

We have generated a He+C+H synthetic white dwarf spectrum with abundances C/He$=0.005$ and
H/He$=10^{-5}$, at the effective temperature and surface gravity shown in Table~\ref{sum2}. We 
have used the latest versions of the plane-parallel, hydrostatic, non-local thermodynamic equilibrium 
(non-LTE) atmosphere and spectral synthesis codes TLUSTY (v200; Hubeny \cite{hubeny88}, Hubeny \& Lanz \cite{hubeny95}) and
SYNSPEC (v48; Hubeny, I. and Lanz, T., private communication). The 
calculation included a full treatment of line blanketing and used state-of-the-art model atoms. In 
brief, the HeII ion incorporated the 19 lowest energy levels, where the dissolution of the higher 
lying levels was treated by means of the occupation probability formalism of Hummer \& Mihalas (\cite{hummer88}),
generalised to the non-LTE situation by Hubeny, Hummer \& Lanz (\cite{hubeny94a}). 
Lines originating from the four lowest levels were treated by means of an approximate 
Stark profile (Hubeny et al. \cite{hubeny94b}) during the calculation of the model structure and in the 
spectral synthesis step. The CIV ion incorporated levels up to n=14 where levels $9\le {\rm n} \le 14$ were
each represented by a superlevel. Where available, photoionization cross sections were obtained from TOPbase.
Transitions between the lowest level and those with n$\le 5$ were represented by Stark profiles. 

The synthetic energy distribution has been normalised to the V magnitude of RE J0503-285 and convolved
with a Gaussian to match the resolution of the SOFI spectrum. This is shown overplotted on the observed 
data in Figure 1. It is worth noting here that our modelling indicates that at the effective temperature
of this white dwarf, the colours V-K, J-H and H-K are rather weak functions of T$_{\rm eff}$. 

\begin{table}
\begin{minipage}{85mm}
\begin{center}
\caption{Summary of additional physical parameters of RE J0503-285 taken from Barstow et al. (1994, 2000). 
Shown in brackets are the values of effective temperature and surface gravity, lying within the uncertainties of
Barstow et al's analysis, which place RE J0503-285 at the largest likely distance. We adopt this latter value 
in setting a limit on the spectral type of a putative cool companion. Note, the distances have been estimated 
using pure-C core evolutionary models of Wood (\cite{wood95}) which exclude a hydrogen layer.}
\label{sum2}
\begin{tabular}{cclcc}
\hline
\hline
 T$_{\rm eff}$(K) & log g & \multicolumn{1}{c}{V} & D(pc) \\
\hline
  72660(75613) & 7.50(7.35) &  $13.90\pm0.01$ &  152(176)  \\
\hline
\end{tabular}
\end{center}

\end{minipage}
\end{table}

\subsection{The search for a cool companion}

We have examined Figure 1 for significant differences in the overall shape or level between the 
observed and synthetic fluxes which can be consistent with the presence of a cool companion. 
Further, we have searched for specific features in the spectrum typical of the energy distributions 
of M or L dwarfs e.g. K~I and Na~I absorption at 1.25$\mu$m and 2.20$\mu$m respectively, CO at 2.3$\mu$m 
and H$_{2}$O centred on 1.15, 1.4 and 1.9$\mu$m. However, no convincing evidence for such has been 
found. Hence, we have added empirical models for low mass stellar and substellar objects, full details 
of which are given in Dobbie et al. (\cite{dobbie05}), to the white dwarf synthetic spectrum and compared these 
composites to the IR data to set an approximate limit on the spectral type of a putative cool companion. 

The fluxes of the empirical models have been scaled to a level appropriate to a location at d=10pc using 
the 2MASS J magnitude of each late-type object and the polynomial fits of Dahn et al. (\cite{dahn02}) and Tinney 
et al. (\cite{tinney03}) to the M$_{\rm J}$ versus spectral type for M6-M9 and L0-L8 field dwarfs/brown dwarfs 
respectively. These fluxes have been further reduced by a factor 1.4, corresponding
to the rms dispersion in the M$_{\rm J}$ versus spectral type relationship of Tinney et al. (\cite{tinney03}). 
Subsequently, the fluxes have been re-calibrated to be consistent with the distance of the DO 
as derived from the measured V magnitude, the effective temperature and theoretical M$_{\rm V}$ and radius from
evolutionary models of pure-C core white dwarfs which include only a He layer mass of $10^{-4}$M$_{\odot}$ (Wood \cite{wood95}). As the effective temperature and surface gravity and hence our distance determination of RE J0503-285 are 
considerably less well constrained than those of the DA white dwarfs we have previously investigated 
(e.g. Dobbie et al. \cite{dobbie05}), for a conservative limit we have adopted the largest distance estimate
consistent with Barstow et al.'s analysis and the error limits given therein (see Table 2). In fact the neglect
of line blanketing arising from metals other than carbon present in the atmosphere of REJ0503-285 impacts on the 
temperature estimate in a such a way as to reinforce our cautious approach (e.g. see Barstow et al. \cite{barstow00}).
Starting with L8 we have progressively added earlier spectral types to the 
synthetic white dwarf spectrum, until it could be concluded with reasonable certainty that the presence 
of a companion of that effective temperature or greater would have been obvious from our data, given the S/N. 


\section{Results}

\subsection{A limit on the mass of a pututive companion to RE J0503-285}

The effective temperatures of very-low-mass stars and substellar objects remain a function of 
both mass and age on the timescale typical of the lifetime of the white dwarfs likely progenitor ($\sim$Gyrs).
Although we can estimate the cooling age of RE J0503-285 using theoretical evolutionary models, as we
don't know with any certainty the mass and hence lifetime of its progenitor star (the initial-mass 
final-mass relationship is poorly constrained for white dwarfs of this mass and in any case may not 
apply to DOs) a robust estimate of the age of a putative associate of the DO is not possible. Therefore,
to use our limit on spectral-type to constrain mass, we instead assume a range of ages broadly 
encompassing the likely value. We assign an approximate effective temperature to the spectral type 
limit shown in Table~\ref{tab2} using the polynomial fit described in Table 4 of Golimowski et al. 
(\cite{golimowski04}). Subsequently, we refer to the low mass stellar/substellar evolutionary models 
for solar metallicity of Baraffe et al. (\cite{baraffe03}), using cubic splines to interpolate between 
their points, to estimate corresponding mass at ages 1Gyr, 5Gyrs and 10Gyrs as shown in Table 3. 

We note that there are a number of examples of post common envelope white dwarf + main sequence binaries,
where a low mass companion has survived a phase of common envelope evolution e.g. RE J0720-318, REJ1016-05,
and RE J2013+400 and now interacts with the white dwarf to influence the composition of its photosphere. 
Our result argues strongly against the presence of an unresolved cool companion to this white dwarf
with M$ \simgreat0.085$M$_{\odot}$ and we can now rule out that the pattern of abundances observed in the 
atmosphere of RE J0503-285 is influenced by ongoing interaction with such an object. However, our result 
does not exclude the possibility that the evolution of RE J0503-285 was affected by the interaction of the 
white dwarf's progenitor star and a close companion (M$\simgreat0.085$M$_{\odot}$) during a common envelope 
phase from which the latter failed to emerge. 
Further, there remains the possibility of a close companion with a mass below our detection threshold.
This may seem unlikely given that an extensive near-IR study of a sample of 371 white dwarfs, summarised
by Farihi, Becklin \& Zuckerman (2005), has revealed only two secondaries with spectral types later than M,  
GD165B (L4; Becklin \& Zuckerman 1998) and GD1400B (L6; Farihi \& Christopher 2004, Dobbie et al. 2005), the 
former at a separation a$\sim$120AU from its white dwarf primary. Nevertheless, there are persistent 
although perhaps controversial claims for L dwarf secondaries in some cataclysmic variables such as EF 
Eri (e.g. Howell \& Ciardi 2001, Harrison et al. 2003).

\begin{table}
\begin{minipage}{85mm}
\begin{center}
\caption{Limiting spectral type, temperature and mass of a putative cool companion to RE J0503-285.
 The effective temperature is estimated from the polynomial relation detailed in Table 4 of Golimowski et 
al. (\cite{golimowski04}). Furthermore, we provide rough upper limits on the mass of a putative cool 
companion as a function of age, by comparing this effective temperature to the predictions of 
the evolutionary models of Baraffe et al. (\cite{baraffe03}).   
 }
\label{tab2}
\begin{tabular}{crrrr}
\hline
SpT  & T$_{\rm eff}$(K) & 1Gyr(M$_{\odot}$) & 5Gyr(M$_{\odot}$) & 10Gyr(M$_{\odot}$) \\
\hline
 M8 & 2500  & 0.084 & 0.085  & 0.085 \\ 

\hline
\end{tabular}
\end{center}
\end{minipage}
\end{table}
  
\subsection{Possible future work}

We point out that RE J0503-285 and the other DO white dwarfs are potential targets for Spitzer imaging to 
search for a mid-IR excess due to companions cooler than spectral type M. Our non-LTE model of REJ 0503-285 
indicates the white dwarf flux at 10$\mu$m to be $\approx$0.04mJy. From the Spitzer spectrum of  DENIS-P
J0255-4700 (Roellig et al. \cite{roellig04}), which we determine to reside at d=5.3pc using the 2MASS magnitude 
(J=13.25$\pm$0.03) and the M$_{\rm J}$ versus spectral type relationship of Tinney et al. (\cite{tinney03}), 
allowing for a 25\% uncertainty in the absolute calibration of the mid-IR data and the uncertainties in late-type 
dwarf fluxes and the white dwarf distance discussed in Section 3.2, we estimate the flux level of an L8 
companion to RE J0503-285 to be $\approx$0.004mJy. As the absolute accuracy of Spitzer photometry is expected
to be $\sim$5\%, it should be possible to extend this search to the regime of the coolest L dwarfs, corresponding
to substellar masses. Further, improved line broadening theories specifically tailored to helium plasmas would 
offer the possibility of reducing the uncertainties on our estimates of the effective temperatures and surface
gravities of DO white dwarfs and allow more robust determinations of their distances.


%

\begin{acknowledgements}
PDD, MRB, ANL and RN are supported by PPARC.
This publication makes use of data products from the Two Micron All Sky Survey, which is a joint project 
of the University of Massachusetts and the Infrared Processing and Analysis Center/California Institute 
of Technology, funded by the National Aeronautics and Space Administration and the National Science Foundation.
We thank the anonymous referee for useful comments which have improved this work.

\end{acknowledgements}

\end{document}